\documentclass[conference]{IEEEtran}
\IEEEoverridecommandlockouts
\usepackage{cite}
\usepackage{amsmath,amssymb,amsfonts}
\usepackage{algorithmic}
\usepackage{graphicx}
\usepackage{textcomp}
\usepackage{xcolor}
\usepackage[linesnumbered,ruled]{algorithm2e}
\def\BibTeX{{\rm B\kern-.05em{\sc i\kern-.025em b}\kern-.08em
    T\kern-.1667em\lower.7ex\hbox{E}\kern-.125emX}}

\usepackage[pscoord]{eso-pic}
\newcommand{\placetextbox}[3]{
    \setbox0=\hbox{#3}
    \AddToShipoutPictureFG*{
    \put(\LenToUnit{#1\paperwidth},\LenToUnit{#2\paperheight}){
    \vtop{{\null}\makebox[0pt][c]{#3}}}
    }
}
\placetextbox{.2}{0.055}{978-3-903176-54-6~\copyright~2023 IFIP}

\newcommand{\Rsym}{\ensuremath{R_{\text{sym}}}}

\newcommand{\SNR}{\ensuremath{\text{SNR}}}
\newcommand{\TXOSNR}{OSNR\textsubscript{TX}}

\newcommand{\rev}[1]{{#1}}

\usepackage{hyperref}
\usepackage{tikz}
\newcommand\copyrighttext{%
  \footnotesize \textcopyright~\the\year~IEEE. Personal use of this material is permitted. Permission from IEEE must be obtained for all other uses, in any current or future media, including reprinting/republishing this material for advertising or promotional purposes, creating new collective works, for resale or redistribution to servers or lists, or reuse of any copyrighted component of this work in other works.
  \href{https://ieeexplore.ieee.org/document/10144887}{E-ISBN: 978-3-903176-54-6}.
  }
\newcommand\copyrightnotice{%
\begin{tikzpicture}[remember picture,overlay]
\node[anchor=north,yshift=-15pt] at (current page.north) {\fbox{\parbox{\dimexpr\textwidth-\fboxsep-\fboxrule\relax}{\copyrighttext}}};
\end{tikzpicture}
\vspace{-0.3cm}
}

\begin{document}
\title{On the Benefits of Rate-Adaptive Transceivers:\\ A Network Planning Study
\thanks{This work has been partially funded in the framework of the CELTIC-NEXT project AI-NET-PROTECT (Project ID C2019/3-4) by the German Federal Ministry of Education and Research (\textbf{\#16KIS1279K}). Carmen Mas-Machuca acknowledges the support of the Federal Ministry of Education and Research of Germany (BMBF) in the programme ``Souverän. Digital. Vernetzt." joint project 6G-life (\textbf{\#16KISK002}).}
}


\author{\IEEEauthorblockN{Jasper M\"uller\IEEEauthorrefmark{1}\IEEEauthorrefmark{2}, Gabriele Di Rosa\IEEEauthorrefmark{1}, Tobias Fehenberger\IEEEauthorrefmark{3}, Mario Wenning\IEEEauthorrefmark{1}\IEEEauthorrefmark{2}, Sai Kireet Patri\IEEEauthorrefmark{1}\IEEEauthorrefmark{2},\\ J\"org-Peter Elbers\IEEEauthorrefmark{1} and Carmen Mas-Machuca\IEEEauthorrefmark{2}} \\
\IEEEauthorblockA{\IEEEauthorrefmark{1} Adtran, Martinsried/Munich, Germany \\
Email: \{jmueller, gdirosa, mwenning, spatri, jelbers\}@adva.com}
\IEEEauthorblockA{\IEEEauthorrefmark{2} Chair of Communication Networks, School of Computation, Information and Technology,\\ Technical University of Munich (TUM), Germany \\
Email: cmas@tum.de}
\IEEEauthorblockA{\IEEEauthorrefmark{3} Adva Network Security, Berlin, Germany\\
Email: tobias.fehenberger@advasecurity.com}}

\maketitle
\copyrightnotice

\begin{abstract}
Flexible-grid Elastic Optical Networks (EONs) have been widely deployed in recent years to support the growing demand for bandwidth-intensive applications. To address this cost-efficiently, optimized utilization of EONs is required. Next-generation bandwidth-variable transceivers (BVTs) will offer increased adaptivity in symbol rate as well as modulation through probabilistic constellation shaping. In this work, we therefore investigate the impact of increased configuration granularity on various aspects of optical networks. We account for practical implementation considerations of BVT configurations for the estimation of the required signal-to-noise ratio. Additionally, an optimization algorithm is presented that selects the most efficient configuration for each considered data rate and bandwidth combination. Based on the advanced transceiver configurations, we conduct a network planning study using a physical-layer-aware algorithm for flexible-grid EONs, and present results for a national and a continental optical backbone network topology. Our research demonstrates that a rise in modulation rate adaptivity results in substantial savings in resources, decreasing the \rev{number} of necessary lightpaths by as much as 20\% in EONs. In contrast, increased symbol rate granularity only results in minor savings.
\end{abstract}


\section{Introduction}
Optical networks are the backbone of today's telecommunications, supporting bandwidth-intensive applications such as high-resolution video streaming, 5G, and autonomous driving. To meet the constantly increasing traffic demands, cost-efficient solutions for capacity scaling are essential.
Flexible-grid Elastic Optical Networks (EONs) have been widely deployed in recent years, thanks to the development of bandwidth-variable transceivers (BVTs).
The increase in the symbol rate per wavelength has been the main driver of reducing the cost per bit and power consumption, leading to the development of coherent optical transceivers that support symbol rates of up to 140~GBd \cite{Jannu}.
However, these large bandwidths also come with highly challenging requirements, particularly for the radio-frequency electronics on the transceiver. It is currently uncertain for how much longer such scaling of the symbol rate will remain technically feasible and economically sensible.

Increasing the maximum data rate configuration of BVTs by increasing the symbol rate primarily improves efficiency for large traffic demands on short-distance links with high signal-to-noise ratio (SNR). Optical long-haul backbone networks often include demands between node-pairs whose distance can vary from less than hundred to several hundred or even thousand kilometers. An efficient utilization of these networks requires a variety of BVT configurations, defined as combinations of symbol rate and modulation format. Advances in rate adaptivity for modulation and symbol rate will allow for fine granularity of configurations in next-generation BVTs \cite{acacia:18}. In this context, the modulation rate adaptivity is achieved through probabilistic constellation shaping (PS) \cite{ChoWinzerPS}.
While these advances in rate adaptivity will enable cost-efficient capacity scaling using existing optical line system (OLS) resources, they also come with additional complexity for planning and operating networks.

Potential benefits of increased rate adaptivity in modulation format using PS have been studied in fixed-grid networks \cite{Mello:14, Ferrari:17}. A potential for significant resource savings has been shown for the use of advanced transmission modes using PS, limited to a channel spacing of 75 or 100~GHz \cite{Karandin:21}. Furthermore, benefits of PS have also been evaluated in multicore fiber networks \cite{Perello:22}, and for online planning in dynamic optical networks \cite{Slovak:18}.
A software defined networking-based optical network planning framework for the adaption of a PS-capable BVT according to optical SNR (OSNR) has been demonstrated in a field trial\cite{Yan:17}.
Several approaches to network planning for flexible-grid EONs have been proposed \cite{SHIRINABKENAR20175, patri2022planning}. The benefits of high-symbol rate next-generation BVTs using traditional modulation formats have also been studied \cite{Pedro:20}.

In this work, we investigate the impact of a further increased configuration granularity on optical networks. We pursue a practical approach to the estimation of the required SNR of PS transmission modes by taking into account a practical forward error correction (FEC) code together with realistic implementation penalties and design choices. PS transmission modes with base constellations of up to 64~QAM, as implemented in commercially available BVTs, are considered.
For further preparation of the network study, an optimization algorithm that selects BVT configurations for a targeted granularity in data rate and bandwidth is presented. For each data rate and bandwidth combination, the algorithm chooses the configuration leading to the lowest required SNR. Finally, a physical-layer-aware network planning algorithm~\cite{patri2022planning} for flexible-grid EONs is used to conduct a network planning study on the Nobel-Germany and the Nobel-EU topologies~\cite{sndlib}.
We show that an increased modulation rate adaptivity leads to significant savings in required number of lighpaths (LPs) of up to 20\%. Furthermore, underprovisioning is reduced by up to 15~percentage points for high requested traffic thereby using the available spectrum more efficiently. An increase in symbol rate granularity on the other hand achieves smaller savings of up to 5\% in LPs while significantly increasing the planning complexity.

\section{Preprocessing of Configuration}
In physical-layer-aware network planning, feasible configurations of a LP are determined by their minimum signal-to-noise ratio (SNR) necessary for error-free transmission (required \SNR). The required \SNR~is determined by the theoretical \SNR~limit given the forward-error-correction overhead ($\text{FEC}_\text{OH}$) as well as implementation penalties which are dependent on the modulation format and the symbol rate. In Section~\ref{sec:reqSNR} we present the calculation of an upper bound of the required SNR for all configurations across different symbol rates and probabilistically shaped modulation formats with variable base constellation size.
In Section~\ref{sec:pre-selection}, the results are used to select only the relevant configurations for a targeted granularity in data rate and symbol rate steps. 

\subsection{Required SNR of Probabilistically Shaped Modulation Formats}\label{sec:reqSNR}
Probabilistically shaped quadrature amplitude modulation (PS~QAM) has the significant advantage of enabling fine information rate adaptation while employing a single FEC code. This advanced signaling technique is usually implemented together with a high-performance soft-decision FEC (SD-FEC) code and the net data rate ($\text{DR}_\text{net}$) is adjusted by tuning the entropy of the transmitted signal, which describes the number of information bits per QAM symbol. Thus, it is offering better performance and finer granularity compared to uniform QAM combined with variable-rate FEC codes \cite{ChoWinzerPS}. This mode of operation requires precise assessment of the achieved $\text{DR}_\text{net}$ and of the related signal quality required at the receiver for each configuration considered for successful network planning. For each configuration, $\text{DR}_\text{net}$ is computed for dual-polarization signals as \cite{ChoWinzerPS}
\begin{align}
\label{InfoRatePSQAM}
\text{DR}_\text{net} = 2\Rsym \left[ H - \log_2(M)\left(1-\frac{1}{1 + \text{FEC}_\text{OH}}\right) \right],
\end{align}
where $M$ is the size of the base QAM constellation, $H$ is the entropy of the signal and \Rsym~is the symbol rate.

Two important observations have to be made when analyzing \eqref{InfoRatePSQAM}: (i) For a fixed non-ideal FEC code, there is a loss in information rate for larger base constellations which is approximately proportional to $\log_2(M)$ \cite{Cho_fixedFEC_PS}. This point, together with the larger implementation penalty associated with higher order modulation, motivates the use of the smallest base constellation that can provide the required entropy as a practical implementation choice. (ii) Increasing the constellation entropy $H$, increases $\text{DR}_\text{net}$, with the upper bound being uniform M-QAM leading to $H = \log_2(M) \rightarrow \text{DR}_\text{net} = 2 \Rsym \log_2(M) / (1 + \text{FEC}_\text{OH})$. 
This results into an effective increase in the exchanged amount of information only if error-free operation is preserved. Clearly, for the same symbol rate, configurations with higher entropy require better received signal quality. 

To predict accurately from pre-FEC data if a received signal can be decoded without errors with a given SD-FEC code, information theoretical metrics such as the normalized generalized mutual information (NGMI) \cite{AlvaradoNGMI} and the asymmetric information (ASI) \cite{ChalmersASI} have been devised. These quantities provide excellent estimators of post-FEC performance irrespective of the modulation format and the $\text{FEC}_\text{OH}$ considered, but have the disadvantage of being less practically translatable into network-level requirements. More conventional and intuitive performance metrics are instead the pre-FEC bit-error-ratio (BER) and the required SNR. The pre-FEC BER is known to be an accurate performance estimator for systems with hard decoding FEC (HD-FEC) but when used to predict the performance of SD-FEC it leads to a variable pre-FEC BER threshold for different modulation formats, with its variability increasing for higher $\text{FEC}_\text{OH}$ \cite{AlvaradoNGMI}. However, the pre-FEC BER has shown to provide quite consistent predictions across PS~QAM formats even for large $\text{FEC}_\text{OH}$ ~$\leq50$\% \cite{ChalmersASI}. 

The solution that we adopt is to consider an established FEC code \cite{adva} with $\text{FEC}_\text{OH}=27\%$, and conservative values for the coding gap ($\approx 0.08$) and the pre-FEC BER threshold ($=3.5\%$). The latter value is certified as a worst case pre-FEC BER that enables successful decoding across the considered modulation formats. Once a given PS~QAM signal distribution is known, it is then possible to translate the pre-BER requirements into conventional SNR requirements assuming additive white Gaussian noise. 
The outcome of this process is a modulation-dependent SNR margin, which is anticipated to have only slight variations across configurations. This considerably simplifies network planning by enabling the use of SNR as the metric for estimating the quality of transmission.
The SNR is calculated, taking into account \TXOSNR, ASE noise from the amplifiers, and interference due to the fiber nonlinearities, computed by the closed-form GN model~\cite{GN}. Additionally, to calculate the required SNR, an implementation penalty for the modulation format of the base constellation is included (1~dB for QPSK, 1.5~dB for 16~QAM and 2~dB for 64~QAM). A second implementation penalty is added as a linear function of the symbol rate, assuming 0.5~dB penalty per 35~GBd, starting with no penalty for a 35~GBd LP. Therefore, it is computed as

\begin{align}
\text{Penalty}_{\Rsym} = 0.5 \text{ dB} \frac{\Rsym - 35 \text{ GBd}}{35 \text{ GBd}}.
\end{align}

\subsection{Pre-selection of Configurations}\label{sec:pre-selection}
We assume that next-generation BVTs will be capable of low-granularity modulation rate adaptivity (0.1 bit/symbol steps between 2 and 6 bit/symbol of entropy) and a tunable symbol rate between 35 and 140 GBd in steps of 2~GBd \cite{acacia:18}. Thereby, 2080 different configurations are possible. The number of useful distinct configurations depends on the targeted data rate granularity of the network operator as well as the spectral slot width in the network. 
Flexible-grid optical networks are usually operated with a spectral slot width of 12.5~GHz while we assume a slot width of 3.125~GHz for next-generation EONs. Furthermore, a data rate granularity of 50~Gbit/s is assumed for the network planning studies while we investigate the effect of three different symbol rate granularities.
We assume a pulse roll-off of 5\% a minimum bandwidth of 37.5~GHz and a maximum bandwidth of 150~GHz. The following rate adaptivity scenarios are investigated:
\begin{itemize}
  \item Scenario \textbf{1)} 4 possible bandwidth configurations of 37.5, 75, 112.5 and 150 GHz,
  \item Scenario \textbf{2)} 10 possible bandwidth configurations with 12.5~GHz resolution,
  \item Scenario \textbf{3)} 37 possible bandwidth configurations with 3.125~GHz resolution.
\end{itemize}
The required \SNR~~($\SNR_{\text{req}}$) is computed as described in Section~\ref{sec:reqSNR}. In order to find the most useful configurations, for each scenario, we start with all combinations of the considered symbol rates and the possible entropies. For each of these configurations, we consider the effective data rate ($\text{DR}_{\text{eff}}$), defined as $\text{DR}_\text{net}$ rounded down to the next 50~GBit/s step.
For each scenario, the relevant configurations are selected such that once a $\text{DR}_{\text{eff}}$ and \Rsym~combination is chosen, the entropy that leads to the lowest $\SNR_{\text{req}}$ is chosen. 
\rev{For a given scenario, we define $\mathcal{D}$ and $\mathcal{S}$ as the set of considered data rates and symbol rates, respectively. $\mathcal{C}_{s, d}$ is the set of configurations with $\text{DR}_\text{eff}=d$ and $\Rsym=s$. The set of considered configurations are chosen by the pre-selection algorithm Alg.~\ref{alg:pre-selection}.
}
\begin{algorithm}
\SetKwInOut{Input}{Input}
\SetKwInOut{Output}{Output}
\Input{Configurations $\mathcal{C}$, considered data rates $\mathcal{D}$ and symbol rates $\mathcal{S}$}
\Output{Pre-selected configurations $\mathcal{C}_{considered}$}
\For{$d \in \mathcal{D}$}{
\For{$s \in \mathcal{S}$}{
$\textbf{select } c \in \mathcal{C}_\text{s,d} \text{ with the lowest } \SNR_{\text{req}}$
$\textbf{add } c \text{ to } \mathcal{C}_{considered}$
}
}
return $\mathcal{C}_{\text{considered}}$   
\caption{Pre-selection Algorithm}
\label{alg:pre-selection}
\end{algorithm}

\begin{figure}[tbp]
\centerline{\includegraphics[width=\columnwidth]{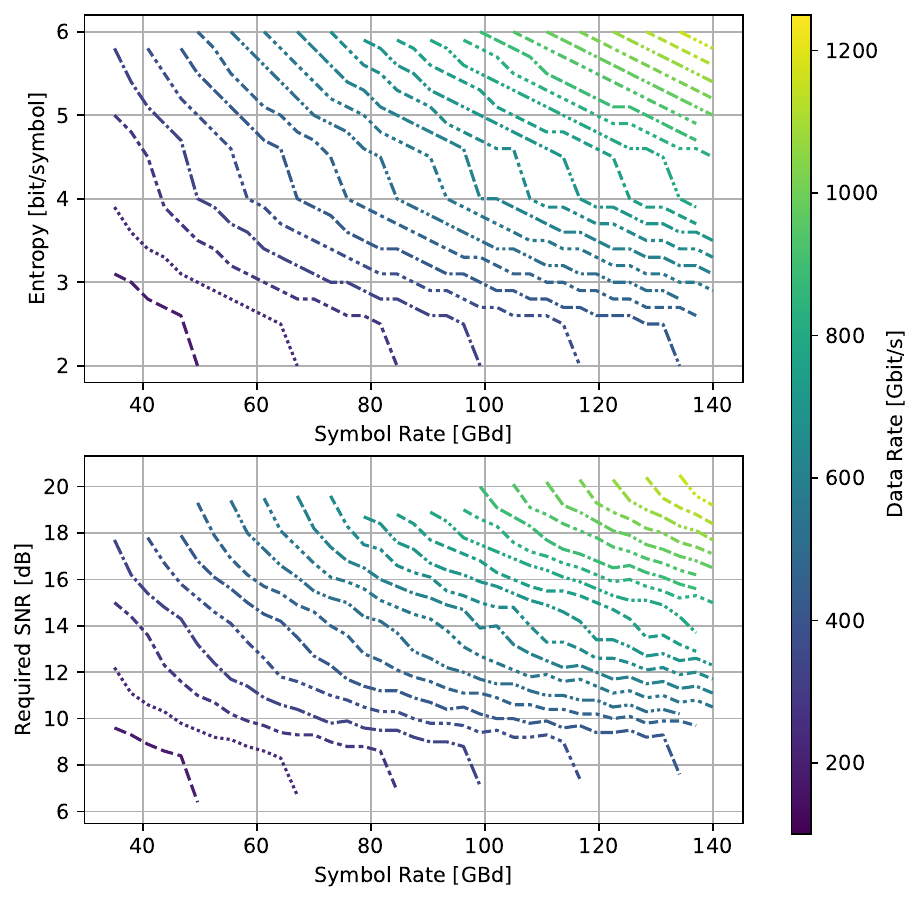}}
\caption{Entropy $H$ (top) and required SNR (bottom) over symbol rate $R_{sym}$ for configurations with 50 Gbit/s data rate granularity and 3.125 GHz channel bandwidth granularity corresponding to Scenario \textbf{3)}.}
\label{fig:configs}
\end{figure}

This method leads to 44, 115 and 423 considered configurations for Scenario \textbf{1)}, \textbf{2)} and \textbf{3)} respectively. Fig.~\ref{fig:configs} shows entropy and required \SNR~over symbol rate for all data rate steps for Scenario \textbf{3)}. We can see that a wide range of symbol rates and required SNR combinations are available for most data rate combinations, enabling highly optimized spectral usage depending on the LPs' \SNR.

\section{Network Planning Study}
In this section, the impact of modulation and symbol rate adaptivity is analyzed in a network planning study on two publicly available network topologies of different characteristics (Tab.~\ref{tab:table-topologies}), representing a national (Nobel-Germany) and a continental (Nobel-EU) backbone network \cite{sndlib}.

\begin{table}[ht!]
\caption{Considered network topologies~\cite{sndlib}.}
   \begin{center}
\begin{tabular}{|c|c|c|c|c|c|}
\hline
\textbf{Topology} & \textbf{Nodes} & \textbf{Links} & \textbf{Demands} & \textbf{Avg. Node} & \textbf{Avg. Path} \\
& & & & \textbf{Degree} & \textbf{Length} \\\hline
Germany      & 17                & 26                & 136        & 3.05   & 420 km     \\ \hline
EU             & 28                & 41          & 378 & 2.92      & 1100 km        \\ \hline
\end{tabular}%
\end{center}
\label{tab:table-topologies}
\end{table}

\begin{figure*}[htbp!]
\centerline{\includegraphics[width=\textwidth]{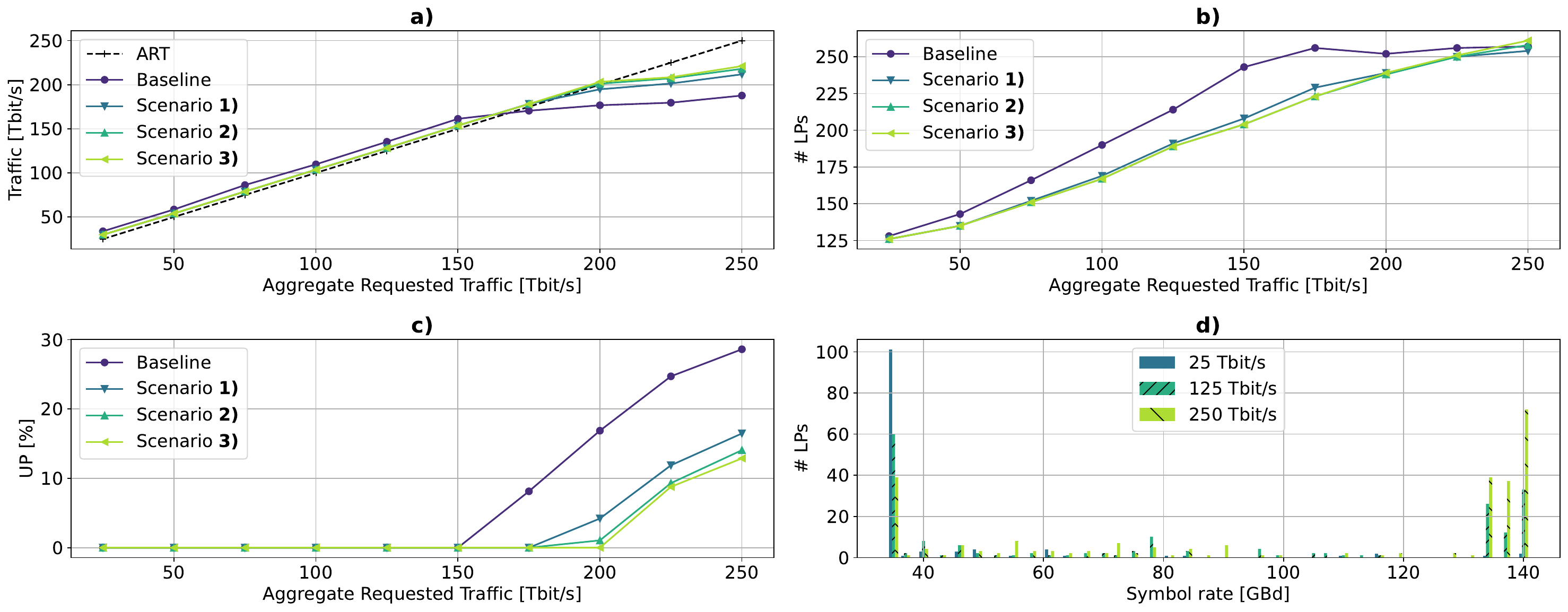}}
\caption{Planning results on Germany: \textbf{a)} provisioned traffic, \textbf{b)} number of deployed LPs, \textbf{c)} underprovisioning, and \textbf{d)} symbol rate distribution for Scenario \textbf{3)}.}
\label{fig:results_GER}
\end{figure*}

\subsection{Setup}
For the planning scenarios, the links are assumed to be standard single-mode fiber (SSMF) links, consisting of 80~km spans with perfect attenuation compensation at the end of each span by an EDFA with 5~dB noise figure. The transmit power spectral density is considered constant for all symbol rates. As a baseline scenario, we consider configurations with symbol rates of 35, 70, 105 and 140 GBd and modulation format of QPSK, 16~QAM and 64~QAM. This baseline is compared to the three scenarios introduced in Section \ref{sec:pre-selection} using PS QAM with different symbol rate granularity. We consider flexible WDM grid networks in the C-Band, with 400 frequency slots of 12.5~GHz each, for the baseline as well as Scenario \textbf{1)} and \textbf{2)}, whereas Scenario \textbf{3)} uses 1600 frequency slots of 3.125~GHz each.

We consider aggregated traffic requests for source-destination node pairs as demands. The demands are weighted according to a traffic model based on the number of data centers and internet exchange points in each ROADM location~\cite{patri2022planning}. To vary the network traffic demands, the individual demands are scaled by the same factor in order to reach different levels of aggregate requested network traffic (ART). The routing, configuration, and spectrum assignment (RCSA) algorithm ~\cite{patri2022planning} considers $k=3$ shortest-path routing. The spectrum is assigned according to the first-fit algorithm. The configurations of each LP are chosen in order to minimize the number of required LPs as primary and the used bandwidth as secondary objective. Candidate configurations are chosen based on the end-of-life (EoL) SNR, assuming a full spectrum with interfering channels of 35 GBd symbol rate. The EoL SNR is computed for each configuration that is valid with respect to the linear \SNR. Configurations with a required SNR threshold lower than the computed SNR are considered. After all LPs are placed, the SNR is recomputed with the now known spectrum of each link to confirm the LP is valid also considering the actual flexible-grid spectrum load \cite{Muller:22}. 
For the analysis of the results of the network planning study, we compare provisioned traffic, number of deployed LPs and underprovisioning ratio (UP) for the different scenarios. UP is defined as \cite{patri2022planning}
\begin{equation}
    \label{eq:up}
    \text{UP} =  \frac{\sum_{{\widetilde{d}}\in \widetilde{D}} \left ( \text{DR}_{\widetilde{d}} - \sum_{{lp} \in \text{LP}_{\widetilde{d}} }\text{DR}_{lp} \right )}{\sum_{{d}\in D} \text{DR}_d },
\end{equation}
where $\text{DR}_{d}$ is the requested traffic of demand $d$, and $\text{DR}_{lp}$ is the data rate of the LP $lp \in \text{LP}_{d}$ provisioned to carry traffic for demand $d$. Here, $\widetilde{D}$ is the subset of demands where the LPs $lp\in \text{LP}_{\widetilde{d}}$ cannot satisfy the requested traffic and it is formally defined as:
\begin{equation}
    \label{eq:up_d_tilde}
    \widetilde{D} = \left \{ \widetilde{d} \in D \mid  \text{DR}_{\widetilde{d}} - \sum_{{lp} \in \text{LP}_{\widetilde{d}} }\text{DR}_{lp} > 0 \right \}.
\end{equation}
Additionally, the number of deployed LPs per symbol rate is compared at different levels of ART for Scenario \textbf{3)}. 

\begin{figure*}[htbp]
\centerline{\includegraphics[width=\textwidth]{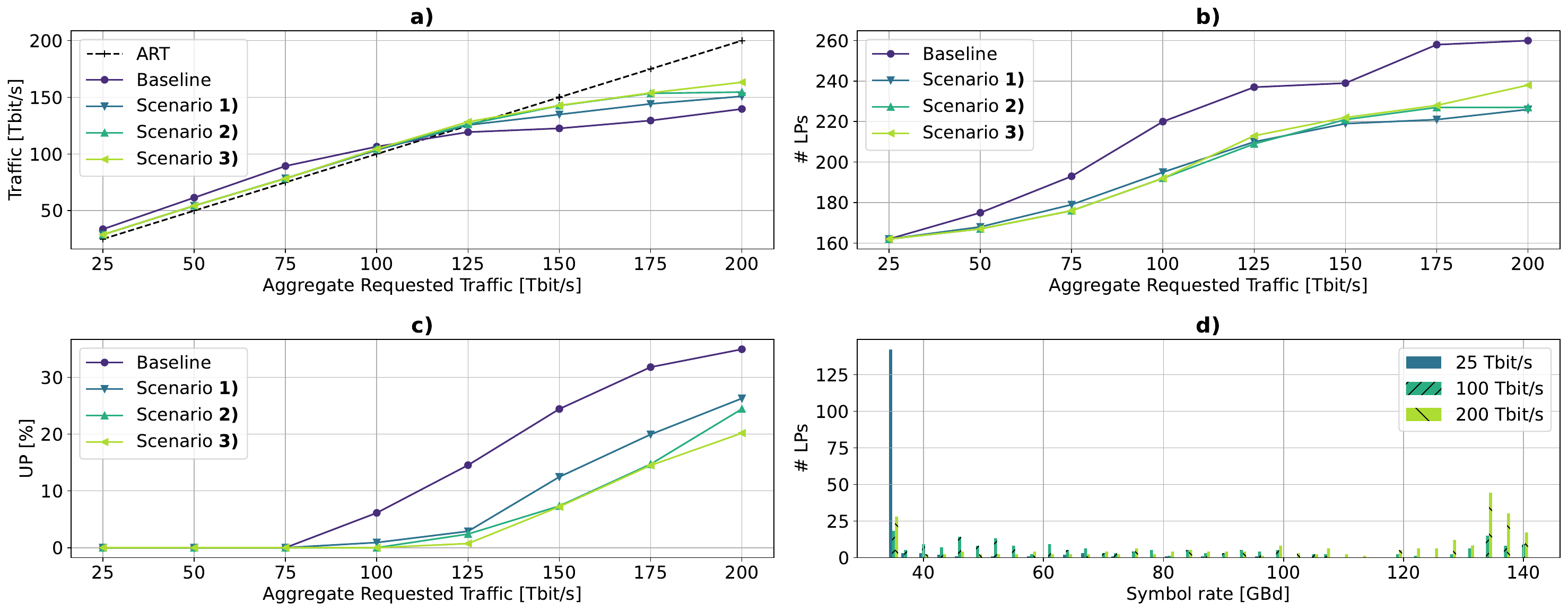}}
\caption{Planning results on EU: \textbf{a)} provisioned traffic, \textbf{b)} number of deployed LPs, \textbf{c)} underprovisioning, and \textbf{d)} symbol rate distribution for Scenario \textbf{3)}.}
\label{fig:results_EU}
\end{figure*}

\begin{figure*}[htbp]
\centerline{\includegraphics[width=\textwidth]{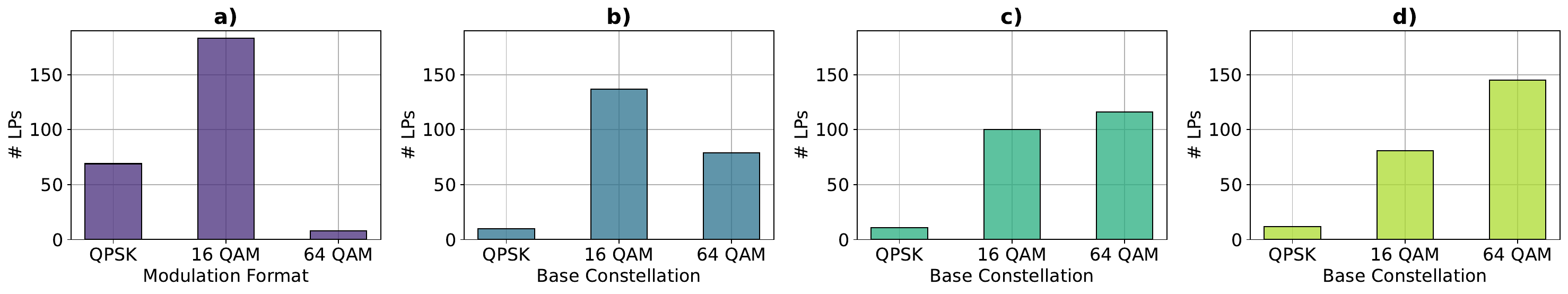}}
\caption{Number of LPs per base constellation on EU for \textbf{a)} the baseline scenario, \textbf{b)} Scenario \textbf{1)} , \textbf{c)}  Scenario \textbf{2)}, and \textbf{d)} Scenario \textbf{3)} at 200~Tbit/s ART.}
\label{fig:modf_EU}
\end{figure*}

\subsection{Results}
The network planning results on the Germany topology are shown in Fig.~\ref{fig:results_GER}. We observe that the baseline scenario shows slightly higher overprovisioning up to 150~Tbit/s of ART when compared to the rate adaptivity scenarios (Fig.~\ref{fig:results_GER}~\textbf{a)}). Overprovisioning occurs because the demanded data rates are on a continuous scale, therefore the next higher feasible data rate will be chosen to fulfill the demand. Since the baseline scenario only considers three modulation format and four symbol rate options some data rates cannot be achieved. Therefore, slightly increased overprovisioning is observed. For high ART, it can be seen that the provisioned traffic of the baseline scenario drops below the ART first. The provisioned traffic of all three rate adaptivity scenarios stays similar over all ART values with Scenario \textbf{2)} and \textbf{3)} provisioning slightly more traffic than Scenario \textbf{1)} for ART above 200~Tbit/s. 

Furthermore, a larger number of LPs is needed to fulfill all demands. At 150~Tbit/s, the baseline scenario requires up to 20\% more LPs than the rate adaptivity scenarios (Fig.~\ref{fig:results_GER} \textbf{b)}). For the rate adaptivity scenarios it can be seen that for ART below 200~Tbit/s Scenario \textbf{1)} requires up to 5\% more LPs than Scenario \textbf{2)} and \textbf{3)} while at the highest considered ART values a slightly lower number of LPs is deployed. Fig.~\ref{fig:results_GER} \textbf{c)} explains this as it can be seen that the UP for Scenario \textbf{1)} is up to 5\% higher than the UP that is observed for Scenario \textbf{2)} and \textbf{3)}. The baseline scenario shows significant higher UP of up to 17~percentage points above the rate adaptivity scenarios for ART above 150~Tbit/s.

These results show a significant benefit when considering rate adaptivity over the baseline scenario as the number of required LPs and UP is noticeably reduced while a higher ART can be provisioned without underprovisioning. The benefit of using a higher symbol rate granularity is relatively minor as at 250~Tbit/s ART the difference in UP between Scenario \textbf{1)} and \textbf{3)} is 4 percentage points while the difference between Scenario \textbf{2)} and \textbf{3)} is only about 1 percentage point. 

Fig.~\ref{fig:results_GER} \textbf{d)} shows the distribution of symbol rates over all LPs for Scenario \textbf{3)} over different levels of ART. It can be seen that for a low level of ART (25~Tbit/s) the minimum considered symbol rate of 35 GBd is chosen while for higher ART, most LP configurations are either at 35 GBd or close to the maximum considered symbol rate of 140 GBd. As the high symbol rate configurations are generally more spectrally efficient due to the absence of guard-bands between multiple channels, these are chosen until one last LP is sufficient to fulfill the remaining requested traffic for a demand. This last LP will be chosen with the minimal possible symbol rate in order to minimize spectral usage. These results explain the lower differences between the rate adaptivity scenarios with different symbol rate granularities when compared to the benefits gained by modulation rate adaptivity, as the minimum and maximum considered symbol rates will be chosen for the majority of the deployed LPs.

As a continental network, the EU topology has a larger number of nodes and demands than the Germany network and more than double the average path length (Tab.~\ref{tab:table-topologies}). The longer paths lead to a lower capacity in this network, considering the used traffic model. The network planning results for the EU network as shown in Fig.~\ref{fig:results_EU} are consistent with our observation on the Germany topology. A large difference between the baseline and the rate adaptivity scenarios is observed as we see more overprovisioning for low ART, until the provisioned traffic drops below ART as the requested traffic cannot be fulfilled at ART values above 75~Tbit/s (Fig.~\ref{fig:results_EU} \textbf{a)}). The baseline scenario requires significantly more LPs than the rate adaptivity scenarios (Fig.~\ref{fig:results_EU} \textbf{b)}) and the UP is up to  17~percentage points higher (Fig.~\ref{fig:results_EU} \textbf{c)}). The differences between the rate adaptivity scenarios are again smaller, although compared to the Germany network a larger gap between Scenario \textbf{2)} and \textbf{3)} is observed for high ART as UP differs by up to 4~percentage points.
Fig.~\ref{fig:results_EU} \textbf{d)} shows that while 35~GBd configurations are used almost exclusively for low ART of 25~Tbit/s in Scenario~\textbf{3)}, for higher ART values the concentration of configurations at the lower and upper end of the considered symbol rates is less pronounced than it is for Germany. The significantly higher number of demands in the EU network lead to a lower requested traffic per node pair on average. Therefore the symbol rate options, that are not at the extremes, are chosen comparably more often for the EU network than for Germany. Therefore, the higher symbol rate granularity of Scenario~\textbf{3)} shows a slightly higher benefit than for Germany when compared to Scenario \textbf{1)} and \textbf{2)}.

In Fig.~\ref{fig:modf_EU} the number of placed LPs for each base constellation is shown at 200~Tbit/s ART on the EU topology for each scenario. We observe that only a few 64~QAM channels are placed in the baseline scenario (Fig.~\ref{fig:modf_EU} \textbf{a)}). In contrast, around one third of the LPs use a 64~QAM base constellation in Scenario \textbf{1)} (Fig.~\ref{fig:modf_EU} \textbf{b)}) going up to around one half and two thirds for Scenario \textbf{2)} (Fig.~\ref{fig:modf_EU} \textbf{c)}) and \textbf{3)} (Fig.~\ref{fig:modf_EU} \textbf{d)}), respectively. As a result, the spectral efficiency is significantly improved compared to the baseline scenario. The results on the Germany topology are consistent with the results shown for EU, with an even higher share of 64~QAM based configurations as the lower average path length enables transmission of signals with higher entropy in the Germany network.

\section{Conclusions}
We investigated the impact of increased configuration granularity on optical networks by calculating the SNR requirements of bandwidth-variable transceiver configurations based on probabilistically shaped modulation formats and taking into account practical implementation aspects for the FEC as well as realistic transceiver implementation penalties. An optimization algorithm that selects the most efficient BVT configurations for a targeted granularity in data rate and bandwidth is presented. Results from a physical-layer-aware network planning study on national and continental optical backbone network topologies show that increased modulation rate adaptivity leads to significant savings in resource usage of up to 20\% in the number of required LPs, making it a cost-efficient solution for capacity scaling in optical networks.

\bibliographystyle{ieeetr}
\bibliography{references.bib}

\begin{thebibliography}{10}

\bibitem{Jannu}
{Acacia}, ``{Acacia Unveils Industry's First Single Carrier 1.2T Multi-Haul
  Pluggable Module}.''
  https://acacia-inc.com/blog/acacia-unveils-industrys-first-single-carrier-1-2t-multi-haul-pluggable-module/,
  2022.

\bibitem{acacia:18}
{Acacia}, ``{{Network Optimization in the 600G Era}}.''
  https://acacia-inc.com/wp-content/uploads/2018/12/Network-Optimization-in-the-600G-Era-WP1218.pdf,
  2018.

\bibitem{ChoWinzerPS}
J.~Cho and P.~J. Winzer, ``Probabilistic constellation shaping for optical
  fiber communications,'' {\em Journal of Lightwave Technology}, vol.~37,
  no.~6, pp.~1590--1607, 2019.

\bibitem{Mello:14}
D.~A.~A. Mello, A.~N. Barreto, T.~C. de~Lima, T.~F. Portela, L.~Beygi, and
  J.~M. Kahn, ``{Optical Networking With Variable-Code-Rate Transceivers},''
  {\em J. Lightwave Technol.}, vol.~32, pp.~257--266, Jan 2014.

\bibitem{Ferrari:17}
A.~Ferrari, M.~Cantono, U.~Waheed, A.~Ahmad, and V.~Curri, ``{Networking
  Benefits of Advanced DSP Techniques and Hybrid Fiber Amplification},'' in
  {\em 2017 19th International Conference on Transparent Optical Networks
  (ICTON)}, pp.~1--4, 2017.

\bibitem{Karandin:21}
O.~Karandin, F.~Musumeci, O.~Ayoub, A.~Ferrari, Y.~Pointurier, and
  M.~Tornatore, ``{Quantifying Resource Savings from Low-Margin Design in
  Optical Networks with Probabilistic Constellation Shaping},'' in {\em 2021
  European Conference on Optical Communication (ECOC)}, pp.~1--4, 2021.

\bibitem{Perello:22}
J.~Perello, J.~M. Gene, and S.~Spadaro, ``{Evaluation of Probabilistic
  Constellation Shaping Performance in Flex Grid over Multicore Fiber Dynamic
  Optical Backbone Networks [Invited]},'' {\em Journal of Optical
  Communications and Networking}, vol.~14, no.~5, pp.~B1--B10, 2022.

\bibitem{Slovak:18}
J.~Slovak, W.~Schairer, M.~Herrmann, K.~Pulverer, and E.~Torrengo, ``{Benefits
  of Performance Awareness in Coherent Dynamic Optical Networks},'' in {\em
  2018 Optical Fiber Communications Conference and Exposition (OFC)}, pp.~1--3,
  2018.

\bibitem{Yan:17}
S.~Yan, F.~N. Khan, A.~Mavromatis, D.~Gkounis, Q.~Fan, F.~Ntavou,
  K.~Nikolovgenis, F.~Meng, E.~H. Salas, C.~Guo, C.~Lu, A.~P.~T. Lau,
  R.~Nejabati, and D.~Simeonidou, ``{Field trial of Machine-Learning-assisted
  and SDN-based Optical Network Planning with Network-Scale Monitoring
  Database},'' in {\em 2017 European Conference on Optical Communication
  (ECOC)}, pp.~1--3, 2017.

\bibitem{SHIRINABKENAR20175}
F.~{Shirin Abkenar} and A.~{Ghaffarpour Rahbar}, ``{Study and Analysis of
  Routing and Spectrum Allocation (RSA) and Routing, Modulation and Spectrum
  Allocation (RMSA) Algorithms in Elastic Optical Networks (EONs)},'' {\em
  Optical Switching and Networking}, vol.~23, pp.~5--39, 2017.

\bibitem{patri2022planning}
S.~K. Patri, A.~Autenrieth, J.-P. Elbers, and C.~Mas-Machuca, ``{Multi-Band
  Transparent Optical Network Planning Strategies for 6G-ready European
  Networks},'' {\em Optical Fiber Technology}, vol.~74, p.~103118, 2022.

\bibitem{Pedro:20}
J.~Pedro, N.~Costa, and S.~Pato, ``{Optical Transport Network Design Beyond 100
  Gbaud [Invited]},'' {\em J. Opt. Commun. Netw.}, vol.~12, no.~2,
  pp.~A123--A134, 2020.

\bibitem{sndlib}
{Zuse Institute Berlin}, ``{SNDlib Problem Instances}.''
\newblock \url{http://sndlib.zib.de/}, Accessed: 2023-01-17.

\bibitem{Cho_fixedFEC_PS}
J.~Cho, S.~L.~I. Olsson, S.~Chandrasekhar, and P.~Winzer, ``{Information Rate
  of Probabilistically Shaped QAM with Non-Ideal Forward Error Correction},''
  in {\em 2018 European Conference on Optical Communication (ECOC)}, pp.~1--3,
  2018.

\bibitem{AlvaradoNGMI}
A.~Alvarado, E.~Agrell, D.~Lavery, R.~Maher, and P.~Bayvel, ``{Replacing the
  Soft-Decision FEC Limit Paradigm in the Design of Optical Communication
  Systems},'' {\em Journal of Lightwave Technology}, vol.~34, no.~2,
  pp.~707--721, 2016.

\bibitem{ChalmersASI}
T.~Yoshida, M.~Karlsson, and E.~Agrell, ``{Performance Metrics for Systems With
  Soft-Decision FEC and Probabilistic Shaping},'' {\em IEEE Photonics
  Technology Letters}, vol.~29, no.~23, pp.~2111--2114, 2017.

\bibitem{adva}
{ADVA}, ``{\textit{TeraFlex\texttrademark}}.''
  https://www.adva.com/en/products/open-optical-transport/fsp-3000-open-terminals/teraflex.
\newblock [Last accessed: January 2023].

\bibitem{GN}
M.~Zefreh, F.~Forghieri, S.~Piciaccia, and P.~Poggiolini, ``{Accurate
  Closed-Form Real-Time EGN Model Formula Leveraging Machine-Learning over 8500
  Thoroughly Randomized Full C-band Systems},'' {\em Journal of Lightwave
  Technology}, vol.~PP, pp.~1--1, 05 2020.

\bibitem{Muller:22}
J.~M\"{u}ller, S.~K. Patri, T.~Fehenberger, H.~Griesser, J.-P. Elbers, and
  C.~Mas-Machuca, ``{QoT Estimation using EGN-assisted Machine Learning for
  Multi-Period Network Planning},'' {\em J. Opt. Commun. Netw.}, vol.~14,
  pp.~1010--1019, Dec 2022.

\end{thebibliography}

\end{document}